\documentclass[10pt,twocolumn,journal]{IEEEtranTCOM}
\normalsize
\usepackage{times}
\usepackage{amsbsy}
\usepackage{amssymb}
\usepackage{fontenc}
\usepackage[usenames]{color}
\usepackage{latexsym}
\usepackage{amsmath}
\usepackage[ansinew]{inputenc}
\usepackage[dvips ]{graphicx}
\input epsf
\usepackage{epic,eepic}
\usepackage{url}
\usepackage{caption}
 \IEEEoverridecommandlockouts

\begin{document}

\title{\huge Low-Complexity Design of Generalized Block Diagonalization Precoding Algorithms for Multiuser MIMO Systems}
\author{\IEEEauthorblockN {Keke Zu, Rodrigo C. de Lamare, {\it Senior Member, IEEE} and Martin Haardt, {\it Senior Member, IEEE}}
\thanks{Part of this work has been submitted to IEEE Asilomar Conference on Signals, Systems and Computers, Pacific Grove, USA, Oct. 2012 \cite{Zu01}}}
\markboth{IEEE Transactions on Communications}%
{Submitted paper}

\maketitle \thispagestyle{empty} \vspace*{-1.5em}

\begin{abstract}
Block diagonalization (BD) based precoding techniques are well-known linear transmit strategies
for multiuser MIMO (MU-MIMO) systems.
By employing BD-type precoding algorithms at the transmit side, the MU-MIMO broadcast channel is decomposed
into multiple independent parallel single user MIMO (SU-MIMO) channels and achieves the
maximum diversity order at high data rates. The main computational complexity of BD-type precoding algorithms comes from
two singular value decomposition (SVD) operations, which depend on the number of users and the dimensions of each user's channel matrix.
In this work, low-complexity precoding algorithms are proposed to
reduce the computational complexity and improve the performance of
BD-type precoding algorithms.
We devise a strategy based on a common channel inversion technique, QR decompositions, and lattice reductions
to decouple the MU-MIMO channel into equivalent SU-MIMO channels. Analytical and simulation results show that the proposed precoding algorithms
can achieve a comparable sum-rate performance as BD-type precoding algorithms, substantial bit error
rate (BER) performance gains, and a simplified receiver structure,
while requiring a much lower complexity.
\end{abstract}


\section{Introduction}
Multiple-input multiple-output (MIMO) systems have drawn a considerable research effort in the past years due to the fact that they
can greatly increase the spectrum efficiency of
wireless communications \cite{Paulraj01,Tse}. In order to meet the continuous growing data traffic, a downlink peak spectrum efficiency of
30 bps/Hz and an uplink peak spectrum efficiency of 15 bps/Hz is proposed in LTE-Advanced \cite{Lte}, and a
configuration of up to $8$ transmit antennas for the downlink is suggested. A new amendment for the WLAN standard IEEE 802.11ac \cite{80211ac} also recommends up to $8$ MIMO spatial streams. Configurations with dozens of antennas are now being considered \cite{chiu}. High-dimensional MIMO systems or large MIMO systems are very promising for the next generation of wireless communication systems due to their potential to improve rate and reliability dramatically \cite{Fredrik}. However, it is a challenge to design a suitable precoding algorithm with good overall
performance and low computational complexity at the same time for high-dimensional MIMO systems.

Unlike the received signal in single user MIMO
(SU-MIMO) systems, the received signals of different users in
multiuser MIMO (MU-MIMO) systems not only suffer
from the noise and the inter-antenna interference but are also affected
by the multiuser interference (MUI). Channel inversion
based precoding or linear precoding algorithms such as zero
forcing (ZF) and minimum mean squared error (MMSE) precoding
\cite{Michael} can still be used to cancel the
MUI, but they result in a reduced throughput or require a higher power
at the transmitter in the MU-MIMO scenarios. As a generalization of the ZF precoding algorithm, block diagonalization (BD) based
precoding algorithms have been proposed in \cite{Spencer02,Choi} for MU-MIMO systems.
However, BD based precoding algorithms only take the MUI into
account and thus suffer a performance loss at low signal to noise ratios
(SNRs) when the noise is the dominant factor. Therefore, a
regularized block diagonalization (RBD) precoding algorithm which introduces a
regularization factor to take the noise term into account has been proposed
in \cite {Veljko}. We term the BD and RBD based precoding schemes as BD-type precoding algorithms in this work for convenience.

The main steps of the BD-type precoding algorithms are
two SVD operations, which need to be implemented for each user. Therefore, the computational complexity of the BD-type precoding algorithms depends on the number of users and the dimensions of each user's channel matrix. For MU-MIMO systems with a large number of users and multiple receive antennas, this could result in a considerable computational cost.
Another distinctive aspect of the BD-type precoding algorithms is that they need a decoding matrix obtained from the second SVD operation to orthogonalize each user's streams. The requirement of this decoding matrix brings extra control overhead or computational complexity \cite{Chae}.

Recent work on BD-type precoding algorithms has focused on how to equivalently implement the BD-type precoding algorithms with less computational complexity. 
A low-complexity generalized ZF channel inversion (GZI) method has
been proposed in \cite{GMI} to equivalently implement the first
SVD operation of the original BD precoding, and a generalized MMSE channel inversion (GMI) method is also developed in \cite{GMI} for the original RBD precoding.
In \cite {Hua} the first SVD operation of the RBD precoding is replaced with a less complex QR
decomposition \cite{MatrixTheo}.
We term the work in \cite{GMI} as GMI-type precoding and the work in \cite{Hua} as QR/SVD-type precoding.
For the second SVD operation, however, both the GMI-type and QR/SVD-type precoding schemes employ it in a similar way as the conventional BD-type precoding algorithms to parallelize each user's streams. Therefore, the second SVD operation needs to be implemented multiple times and the decoding matrix for the effective channel still needs to be known or estimated at the receiver of each user for the GMI-type or QR/SVD-type precoding algorithms.

The GMI-type and QR/SVD-type techniques are solely low complexity equivalent implementations of the BD-type precoding algorithms.
As an improvement of the BD-type precoding algorithms, a low-complexity lattice reduction-aided RBD (LC-RBD-LR) type precoding algorithms has been proposed in \cite{Zu03, Zu02} based on the QR decomposition scheme. Not only much less complexity but also considerable BER gains are achieved by the LC-RBD-LR-type precoding algorithms.
However, the QR decomposition in LC-RBD-LR-type precoding algorithms still needs to be implemented for each user, which could result in a high complexity for large MIMO systems.

A new category of low-complexity high performance precoding algorithms based on the channel inversion scheme is proposed in this work. 
A simplified GMI (S-GMI) precoding scheme which employs a common channel inversion for all users is developed first. Equivalent parallel SU-MIMO channels are obtained from the S-GMI precoding process. Then, these effective channels are transformed into the lattice space by utilizing the lattice reduction (LR) technique \cite{LR}, whose complexity is dictated by a QR decomposition. Linear precoding strategies are applied in the lattice space to parallelize each user's streams.
Finally, the proposed lattice reduction-aided simplified GMI (LR-S-GMI) precoding algorithms are obtained. According to the specific linear precoding constraint used, the proposed LR-S-GMI-type precoding algorithms are categorized as LR-S-GMI-ZF and LR-S-GMI-MMSE, respectively.

The algorithm structure of the proposed LR-S-GMI-type precoding is different from the LC-RBD-LR-type precoding since the channel inversion is only implemented once for all users, while the QR decomposition needs to be implemented multiple times in LC-RBD-LR-type precoding. Therefore, the computational complexity can be reduced considerably by the proposed LR-S-GMI-type precoding.
A comprehensive mathematical analysis is developed to analyze and predict the performance of the proposed LR-S-GMI-type precoding algorithms. The simulation results verified that the proposed LR-S-GMI-type precoding algorithms have the lowest computational complexity compared to BD-type \cite{Spencer02, Veljko}, GMI-type \cite{GMI}, QR/SVD-type \cite{Hua} and LC-RBD-LR-type \cite{Zu03} precoding algorithms, a comparable sum-rate performance as BD-type precoding algorithms, and substantial BER performance gains over prior art.

The main contributions of the work are summarized below:
\begin{enumerate}
   \item A simplified GMI (S-GMI) precoding is developed in this work as an improvement of the original RBD in \cite{Veljko}. A mathematical analysis is given to show that the S-GMI has a better BER performance and much less complexity than that of RBD, which is a clear difference compared to the GMI in \cite{GMI} which only provides an equivalent implementation of RBD.
   \item A new category of low-complexity high-performance LR-S-GMI-type precoding algorithms is proposed for MU-MIMO systems based on a channel inversion technique, QR decompositions, and lattice reductions.
   \item The BD-type precoding algorithms are systematically analyzed and summarized. We show that the computational complexity of the BD-type precoding depends on the number of users and the system dimensions.
   \item A comprehensive performance analysis is carried out in terms of BER performance, achievable sum-rate, and computational complexity.
   \item A simulation study of the proposed algorithms under imperfect channel situations is also conducted, which completes this paper.
 \end{enumerate}

The proposed and existing precoding techniques are all performed with the help of downlink channel state information (CSI). The assumption that full CSI is available at the transmit side is valid in time-division duplex (TDD) systems because the uplink and downlink share the same frequency band. For frequency-division duplex (FDD) systems, however, the CSI needs to be estimated at the receiver and fed back to the transmitter.

This paper is organized as follows. The system model is given in Section II. A brief review of the BD-type precoding algorithms is presented in Section III. The proposed LR-S-GMI-type precoding algorithms are described in detail in Section IV and the performance analysis is developed in Section V. Simulation results and conclusions are displayed in Section VI and Section VII, respectively.

{\it Notation}: Matrices and vectors are denoted by upper and lowercase boldface letters, and the transpose, Hermitian transpose, inverse, pseudo-inverse of a matrix $\boldsymbol B$ are described by $\boldsymbol B^T$, $\boldsymbol B^H$, $\boldsymbol B^{-1}$, $\boldsymbol B^{\dagger}$, respectively. The trace, determinant, Frobenius norm, round function are denoted as $Tr(\cdot)$, $det(\cdot)$, $\|\cdot\|_{\rm F}$, $\lceil\cdot \rfloor$. With ${\rm diag}\{\boldsymbol B_1,\boldsymbol B_2,\ldots,\boldsymbol B_K$\} creates a block diagonal matrix with the matrices $\boldsymbol{B}_k$ on the main diagonal.

\section{System Model}
  We consider an uncoded MU-MIMO downlink channel, with $N_T$ transmit antennas
at the base station (BS) and $N_i$ receive antennas at the $i$th
user equipment (UE). With $K$ users in the system, the total number
of receive antennas is $N_R=\sum _{i=1}^{K}N_i$. A block diagram of such a system is illustrated in Fig. 1.
\begin{figure}[htp]
\begin{center}
\def\epsfsize#1#2{0.95\columnwidth}
\epsfbox{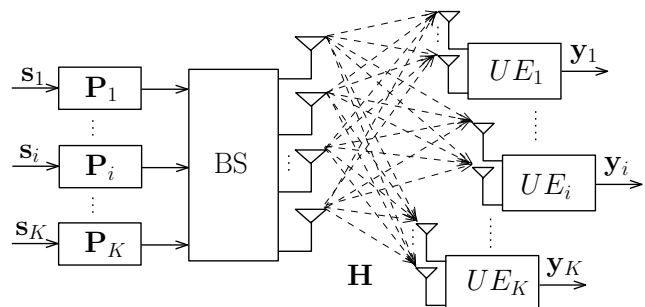}
 \vspace{-0.5em}
\caption{MU-MIMO System Model}
\end{center}
\end{figure}

From the system model, the combined channel matrix $\boldsymbol H$ and the joint precoding matrix $\boldsymbol P$ are given by
 \begin{align}
&\boldsymbol H=[\boldsymbol H_1^T ~ \boldsymbol H_2^T ~ \ldots ~
\boldsymbol H_K^T]^T\in\mathbb{C}^{N_R\times N_T},\\
&\boldsymbol P=[\boldsymbol P_1 ~ \boldsymbol P_2 ~ \ldots ~
\boldsymbol P_K]\in\mathbb{C}^{N_T\times N_R},
\end{align}
where ${\boldsymbol H_i}\in\mathbb{C}^{N_i\times N_T}$ is the $i$th user's channel matrix.
The quantity ${\boldsymbol P_i}\in\mathbb{C}^{N_T\times N_i}$ is the $i$th user's precoding
matrix. We assume a flat
fading MIMO channel and the received signal $\boldsymbol y_i\in\mathbb{C}^{N_i}$ at the $i$th user is given by
  \begin{align}
  \boldsymbol y_i =\boldsymbol H_i\boldsymbol x_i+\boldsymbol H_i\sum _{j=1,j\neq i}^{K}\boldsymbol x_j + {\boldsymbol n}_i,
  \end{align}
  where the quantity ${\boldsymbol x_i}\in\mathbb{C}^{N_i}$ is the
$i$th user's transmitted signal, and ${\boldsymbol
n}_i\in\mathbb{C}^{N_i} $ is the $i$th user's Gaussian noise with
independent and identically distributed (i.i.d.)
entries of zero mean and variance $\sigma_n^2$. Assuming that the average
transmit power for each user is $\xi_i$, then, the power constraint ${\rm E}\| \boldsymbol x_i\|^2=\xi_i$ is imposed. We construct
an unnormalized signal $\boldsymbol s_i$ such that
\begin{align}
\boldsymbol x_i={\boldsymbol s_i \over \sqrt {{\rm E}_{\gamma_i}}},
\end{align}
where $\boldsymbol s_i=\boldsymbol P_i\boldsymbol d_i$ with $\boldsymbol d_i$ being the transmit data vector and ${\rm E}_{\gamma_i}$ is the average energy of $\gamma_i$ with $\gamma_i=\|\boldsymbol s_i\|^2/\xi_i$. The physical meaning of dividing $\boldsymbol s_i$ by the scalar $\sqrt {{\rm E}_{\gamma_i}}$ is to make sure the average transmit power $\xi_i$ is still the same after the precoding process. With this normalization, $\boldsymbol x_i$ obeys ${\rm E}\| \boldsymbol x_i\|^2=\xi_i$.

The received signal $\boldsymbol y_i$ is weighted by the scalar $\sqrt {{\rm E}_{\gamma_i}}$ to form the estimate
  \begin{align}
  \boldsymbol {\hat d}_i =\sqrt {{\rm E}_{\gamma_i}}\boldsymbol y_i. 
  \end{align}
Note that it is necessary to cancel $\sqrt {{\rm E}_{\gamma_i}}$ out at the receiver to get the correct amplitude of the desired signal part.
The average energy ${\rm E}_{\gamma_i}$ is independent from the channel and the data, which means the receivers do not need to know the instantaneous CSI for the precoding techniques to work \cite{Peel}. As analyzed and illustrated in \cite{Hochwald}, however, the performance difference between the average ${\rm E}_{\gamma_i}$ and the instantaneous $\gamma_i$ is very small. Therefore, we follow the strategy developed in \cite{Peel} and \cite{Hochwald} to assume the receivers need to know only $\sqrt {{\rm E}_{\gamma_i}}$ but use $\gamma_i$ instead of ${\rm E}_{\gamma_i}$ for simulation convenience as ${\gamma_i}$ is simpler to compute. The simulation results represent the performance of either normalization method. In this case, we can replace (4) and (5) with the instantaneous $\gamma_i$ in the simulations and employ
\begin{align}
\boldsymbol x_i={\boldsymbol s_i \over \sqrt {\gamma_i}}~{\rm and}~ \boldsymbol {\hat d}_i =\sqrt \gamma_i \boldsymbol y_i.
\end{align}

\section{Review of BD-type Precoding Algorithms}
The design of BD-type precoding algorithms is performed in two
steps \cite{Spencer02, Veljko}. The first precoding filter is used to completely eliminate (by BD) or balance the MUI with noise (by RBD), then exact (by BD) or approximate (by RBD) parallel SU-MIMO channels are obtained. The second precoding filter is implemented to parallelize each user's streams.
Correspondingly, the precoding matrix $\boldsymbol P_i$ for the $i$th user can be rewritten in two parts as
\begin{align}
\boldsymbol P_i= {\boldsymbol P}_i^a{\boldsymbol P}_i^b,
\end{align}
where ${\boldsymbol P}_i^a\in\mathbb{C}^{N_T\times M_i}$ and ${\boldsymbol P}_i^b\in\mathbb{C}^{M_i\times N_i}$. The parameter $M_i$ is dependent on
the specific choice of the precoding algorithm.
We exclude the $i$th user's channel matrix
and define $\boldsymbol{\overline H}_i$ as
\begin{align}
\boldsymbol{\overline H}_i=[\boldsymbol H_1^T ~\dots~\boldsymbol H_{i-1}^T ~ \boldsymbol H_{i+1}^T ~ \ldots ~ \boldsymbol
H_K^T]^T\in\mathbb{C}^{\overline N_i\times N_T},
\end{align}
where $\overline N_i=N_R-N_i$. Then, the interference generated to the other users is determined by $\boldsymbol{\overline H}_i\boldsymbol P_i^a$.

In order to eliminate all the MUI, we impose the constraint that
\begin{align}
\forall i\in(1,\ldots,K)~\boldsymbol{\overline H}_i\boldsymbol P_i^a = \boldsymbol 0~{\rm s.t.}~\rm E\|\boldsymbol x_i\|^2=\xi_i.
\end{align}
We term (9) as the BD constraint. Note that the BD constraint is actually an extension of the ZF constraint in \cite{Michael} for MU-MIMO with multiple receive antennas.
In order to take the noise term into account as well, an RBD constraint is developed in \cite{Veljko} and given by
\begin{align}
\boldsymbol P_i^a = \min_{\boldsymbol P_i^a} \rm E\lbrace\|\boldsymbol{\overline H}_i\boldsymbol P_i^a\|^2+{\gamma_i\|\boldsymbol n_i\|^2}\rbrace \nonumber \\
{\rm s.t.}~\rm E\|\boldsymbol x_i\|^2=\xi_i.
\end{align}

Assuming that the rank of $\boldsymbol{\overline H}_i$ is $\overline L_i$, define the SVD of $\boldsymbol{\overline H}_i$
\begin{align}
\boldsymbol{\overline H}_i=\boldsymbol{\overline U}_i\boldsymbol{\overline \Sigma}_i\boldsymbol{\overline V}_i^H=\boldsymbol{\overline U}_i\boldsymbol{\overline \Sigma}_i[\begin {array}{c c} \boldsymbol{\overline V}_i^{(1)} & \boldsymbol{\overline V}_i^{(0)}\end {array}]^H,
\end{align}
where $\boldsymbol{\overline U}_i\in\mathbb{C}^{\overline N_i\times \overline N_i}$ and $\boldsymbol{\overline V}_i\in\mathbb{C}^{N_T\times N_T}$ are unitary matrices. The diagonal matrix $\boldsymbol{\overline \Sigma}_i \in\mathbb{C}^{\overline N_i\times N_T}$ contains the singular
values of the matrix $\boldsymbol{\overline H}_i$. Factorizing $\boldsymbol{\overline V}_i$ into two parts, $\boldsymbol{\overline V}_i^{(1)}\in\mathbb{C}^{N_T\times\overline L_i}$ consists of the first $\overline L_i$ non-zero singular vectors and $\boldsymbol{\overline V}_i^{(0)}\in\mathbb{C}^{N_T\times {(N_T-\overline L_i)}}$ holds the last $N_T-\overline L_i$ zero singular vectors. Thus, $\boldsymbol{\overline V}_i^{(0)}$ forms an orthogonal basis for the null space of $\boldsymbol{\overline H}_i$. The solution for the BD constraint (9) is given by
\begin{align}
{\boldsymbol P_i^a}^{\rm (BD)}=\boldsymbol{\overline V}_i^{(0)}.
\end{align}
As shown in \cite{Veljko}, the solution for the RBD constraint can be obtained as
\begin{align}
{\boldsymbol P_i^a}^{\rm (RBD)}=\boldsymbol{\overline V}_i(\boldsymbol{\overline \Sigma}_i^T\boldsymbol{\overline \Sigma}_i+\alpha\boldsymbol I_{N_T})^{-1/2},
\end{align}
where $\alpha={N_R\sigma_n^2\over \xi}$ is the regularization factor with $\xi$ is the whole average transmit power.

After the first precoding process, the MU-MIMO channel is decoupled into a set of $K$ parallel independent SU-MIMO channels by the BD precoding. For the RBD precoding, there are residual interferences between these channels due to the regularization process, but, these interferences tend to zero at high SNRs.
Therefore, the effective channel matrix for the $i$th user can be expressed as
\begin{align}
\boldsymbol H_{{\rm eff}_i}=\boldsymbol H_i\boldsymbol P_i^a.
\end{align}
Define $L_{\rm eff}={\rm rank}(\boldsymbol H_{{\rm eff}_i})$ and consider the second SVD operation on the effective channel matrix
\begin{align}
\boldsymbol H_{{\rm eff}_i}=\boldsymbol{U}_i\boldsymbol{\Sigma}_i{\boldsymbol{V}_i}^H=\boldsymbol{U}_i{\left [\begin {array}{c c} \boldsymbol \Sigma_i & 0\\ 0 & 0\end {array} \right]}[\begin {array}{c c} \boldsymbol{V}_i^{(1)}& \boldsymbol{V}_i^{(0)} \end {array}]^H,
\end{align}
using the unitary matrix $\boldsymbol{U}_i \in\mathbb{C}^{L_{\rm eff}\times L_{\rm eff}}$ and $\boldsymbol{V}_i^{(1)}$ 
contains the first $L_{\rm eff}$ singular vectors. The second precoding filters for BD and RBD precoding can be respectively obtained as
\begin {align}
{\boldsymbol P_i^b}^{\rm (BD)}=\boldsymbol{V}_i^{(1)}{\boldsymbol\Lambda}^{\rm (BD)},\\
{\boldsymbol P_i^b}^{\rm (RBD)}=\boldsymbol{V}_i{\boldsymbol\Lambda}^{\rm (RBD)},
\end {align}
where ${\boldsymbol\Lambda}$ is the power loading matrix that depends on the optimization criterion. An example power loading is the water filling (WF) \cite{Paulraj01}. The $i$th user's decoding matrix is obtained as
\begin {align}
{\boldsymbol G_i}=\boldsymbol{U}_i^H,
\end {align}
which needs to be known by each user's receiver.

Note that for the conventional BD-type precoding algorithms, there is a dimensionality constraint below to be satisfied
\begin {align}
N_T>{\rm max}\{{\rm rank}(\boldsymbol{\overline H}_1),{\rm rank}(\boldsymbol{\overline H}_2),\ldots,{\rm rank}(\boldsymbol{\overline H}_K)\}.
\end {align}
Then, we can get the matrix dimension relationship as $N_T\geq N_R>\overline N_i\geq\overline L_i>N_i\geq L_{\rm eff}$. Note that the first SVD operation in (11) needs to be implemented $K$ times on $\boldsymbol{\overline H}_i$ with dimension $\overline N_i\times N_T$ and the second SVD operation in (15) needs to be implemented $K$ times on $\boldsymbol H_{{\rm eff}_i}$ with dimensions $L_{\rm eff}\times {(N_T-\overline L_i)}$ for the BD precoding and $L_{\rm eff}\times N_T$ for the RBD precoding.
From the above analysis, most of the computational complexity of the BD-type precoding algorithms comes from the two SVD operations which make the computational complexity of the BD-type precoding algorithms increase with the number of users $K$ and the system dimensions.

\section{Proposed S-GMI Based Precoding Algorithms}
In this section, we describe the proposed LR-S-GMI-type precoding algorithms based on a strategy that employs a
channel inversion method, QR decompositions, and lattice reductions.  Similar to the BD-type precoding algorithms, the design of the proposed LR-S-GMI-type precoding algorithms is computed in two steps.

First, we obtain the first precoding filter $\boldsymbol P_i^a$
for the LR-S-GMI-type precoding algorithms by using one channel inversion and $K$ QR decompositions each implemented on individual users with matrix dimension $N_i\times N_i$.
By applying the MMSE inversion to the combined channel matrix, we have
\begin{equation}
\begin{split}
\boldsymbol H^{\dag}_{{\rm mse}} & =\boldsymbol H^H(\boldsymbol H\boldsymbol
H^H+\alpha \boldsymbol I)^{-1} \\ & =[\boldsymbol
H_{1,{\rm mse}},\boldsymbol H_{2,{\rm mse}},\ldots,\boldsymbol
H_{K,{\rm mse}}].
\end{split}
\end{equation}
where $\boldsymbol H_{i,{\rm mse}}\in\mathbb{C}^{N_T\times\ N_i}$ is the sub-matrix of $\boldsymbol H^{\dag}_{{\rm mse}}$. Considering a high SNR case,
it can be shown that the regularization factor $\alpha$ approaches zero and thus we have ${\boldsymbol H} {\boldsymbol H}^\dag_{{\rm mse}} \approx \boldsymbol I_{N_T}$ \cite{Michael}.
This means the off-diagonal block matrices of ${\boldsymbol H} {\boldsymbol H}^\dag_{{\rm mse}}$ converge to zeros with the increase of SNR.
Hence, the matrix ${\boldsymbol H}_{i,{\rm mse}}$ is approximately in the null space of ${\boldsymbol{\overline H}}_i$ defined in (8), that is,
\begin{align}
{\boldsymbol{\overline H}}_i
{\boldsymbol H}_{i,{\rm mse}}\approx \boldsymbol 0.
\end {align}
Considering the QR decomposition of $\boldsymbol H_{i,{\rm mse}}=\boldsymbol Q_{i,{\rm mse}} \boldsymbol R_{i,{\rm mse}}$, we have
\begin{align}
\boldsymbol{\overline H}_i\boldsymbol H_{i,{\rm mse}} =
\boldsymbol{\overline H}_i\boldsymbol Q_{i,{\rm mse}} \boldsymbol
R_{i,{\rm mse}} \approx \boldsymbol 0 ~{\rm for}~i=1,\ldots,K,
\end{align} where ${\boldsymbol Q}_{i,{\rm mse}}\in\mathbb{C}^{N_T\times N_i}$ is an orthogonal
matrix and $\boldsymbol R_{i,{\rm
mse}}\in\mathbb{C}^{N_i\times N_i}$ is an upper triangular matrix.
Since $\boldsymbol R_{i,{\rm mse}}$ is invertible, we have
\begin{align}
\boldsymbol{\overline H}_i\boldsymbol Q_{i,{\rm
mse}}\approx \boldsymbol 0.
\end {align}
Thus, $\boldsymbol Q_{i,{\rm mse}}$
satisfies the RBD constraint (10) to balance the MUI and the noise term.

We have simplified the design of the first precoding filter $\boldsymbol P_i^a$ here as compared to \cite{GMI} where a
residual interference suppression filter $\boldsymbol T_i$ is
applied after the first precoding process $\boldsymbol P_i^a$. The filter $\boldsymbol T_i$ increases the complexity and
cannot completely cancel the MUI. Therefore, we omit the residual
interference suppression part since it is not necessary for the RBD constraint based
precoding. We term the simplified GMI as S-GMI in this work.
Then, the first precoding filter for S-GMI can be obtained as
\begin{align}
\boldsymbol P_i^a=\boldsymbol Q_{i,{\rm mse}},
\end{align}
where $\boldsymbol P_i^a\in\mathbb{C}^{N_T\times N_i}$. By implementing the QR decomposition in (22) $K$ times on ${\boldsymbol H}_{i,{\rm mse}}$ with dimension $N_i\times N_i$, the first combined precoding matrix for S-GMI is
\begin{align}
\boldsymbol P^a=[\boldsymbol P_1^a,~\boldsymbol P_2^a,~\ldots,~\boldsymbol P_K^a].
\end{align}
Note the $K$ QR decompositions of the LC-RBD-LR-type precoding in \cite{Zu03, Zu02} are implemented on $\boldsymbol{\overline H}_i$ with dimension $\overline N_i\times N_i$.
The S-GMI algorithm can be completed by applying the SVD operation to the effective channel matrix $\boldsymbol H_{{\rm eff}_i}=\boldsymbol H_i\boldsymbol P_i^a=\boldsymbol{U}_i\boldsymbol{\Sigma}_i{\boldsymbol{V}_i}^H$. Then, the second precoding filter of S-GMI is obtained as ${\boldsymbol P_i^b}=\boldsymbol{V}_i$. The S-GMI algorithm is summarized in Table I.

\begin{table}[!t]
\caption{The S-GMI Precoding Algorithm} 
\centering 
\begin{tabular}{l l} 
\hline\hline 
Steps & Operations \\ [0.5ex] 
\hline 
& {\bf Applying the MMSE Channel Inversion}~~~~~~~~~~~~~~~~~~~~~\\
(1)& $\boldsymbol H^{\dag}_{{\rm mse}}  =(\boldsymbol H^H\boldsymbol
H+\alpha \boldsymbol I)^{-1}\boldsymbol H^H$\\
(2)& for i~=~1~:~$ K$\\
(3)&~~~~~$[\boldsymbol Q^\dag_{i,{\rm mse}}~ {\boldsymbol R^\dag_{i,{\rm mse}}}]={\rm QR}(\boldsymbol H^\dag_{i,{\rm mse}},~0)$\\
(4)&~~~~~$\boldsymbol P_i^a=\boldsymbol Q^\dag_{i,{\rm mse}}$\\
(5)&~~~~~$\boldsymbol H_{{\rm eff}_i}=\boldsymbol H_i\boldsymbol P_i^a=\boldsymbol{U}_i\boldsymbol{\Sigma}_i{\boldsymbol{V}_i}^H$\\
(6)&~~~~~$ {\boldsymbol P}_i^b=\boldsymbol{V}_i$\\
(7)&~~~~~$ {\boldsymbol G}_i=\boldsymbol{U}_i^H$\\
(8)&end\\
& {\bf Compute the overall precoding and decoding matrix}\\
(9)&$\boldsymbol P^a=[\boldsymbol P_1^a,~\boldsymbol P_2^a,~\ldots,~\boldsymbol P_K^a]$\\
(10)&$\boldsymbol P^b={\rm diag}\{\boldsymbol P_1^b,\boldsymbol P_2^b,\ldots,\boldsymbol P_K^b\}$\\
(11)&$\boldsymbol P=\boldsymbol P^a\boldsymbol P^b$~~~$\boldsymbol G={\rm diag}\{{\boldsymbol G}_1,~{\boldsymbol G}_2,~\ldots,~{\boldsymbol G}_K\}$\\
& {\bf Calculate the scaling factor $\gamma$}\\
(12)&$\gamma=(\|\boldsymbol P\boldsymbol d\|_F^2/E_s)$\\
& {\bf Get the received signal}\\
(13)& $\boldsymbol y=\boldsymbol G(\boldsymbol H\boldsymbol P\boldsymbol d+\sqrt\gamma\boldsymbol n)$\\
[1ex] 
\hline 
\end{tabular}
\end{table}

Similarly, the extension of the channel inversion method from the RBD constraint based precoding to the BD constraint based precoding is straightforward on
\begin{align}
\boldsymbol H^{\dag}_{\rm {zf}} =\boldsymbol H^H(\boldsymbol H\boldsymbol H^H)^{-1}=[\boldsymbol H_{1,\rm {zf}},\boldsymbol H_{2,\rm {zf}},\ldots,\boldsymbol H_{K,\rm {zf}}].
\end{align}
Moreover, the obtained MUI is strictly zero as $\boldsymbol{\overline H}_i\boldsymbol H_{i,\rm {zf}}=\boldsymbol 0$. Assuming the QR decomposition of $\boldsymbol H_{i,\rm {zf}}$ is $\boldsymbol H_{i,\rm {zf}}=\boldsymbol Q_{i,\rm {zf}}\boldsymbol R_{i,\rm {zf}}$, then, we have
\begin{align}
\boldsymbol{\overline H}_i\boldsymbol Q_{i,\rm {zf}}=\boldsymbol 0.
\end{align}
Thus, $\boldsymbol Q_{i,\rm {zf}}$ satisfies the BD constraint (9). The first precoding matrix for the BD constraint based precoding can be equivalently obtained as
\begin{align}
\boldsymbol P_i^a=\boldsymbol Q_{i,\rm {zf}}.
\end{align}
This equivalent method is termed as GZI in \cite{GMI}.

For the proposed LR-S-GMI-type precoding algorithms, we get the first precoding filter as S-GMI in (24), while we employ the LR-aided linear precoding technique instead of the SVD operation in S-GMI to obtain the second precoding filter $\boldsymbol P_i^b$.
The aim of the LR transformation is to find a new basis
$\boldsymbol {\tilde H}$ which is nearly orthogonal compared to the
original matrix $\boldsymbol H$ for a given lattice $L(\boldsymbol
H)$. The most commonly used LR algorithm has been first proposed by
Lenstra, Lenstra and L. Lov\'asz (LLL) in \cite{LLL} with polynomial
time complexity. In order to reduce the
computational complexity, a complex LLL (CLLL)
algorithm was proposed in \cite{CLLL}, which reduces the overall complexity of the LLL algorithm by nearly half without
sacrificing any performance. We employ the CLLL algorithm to implement the LR transformation in this work.

After the first precoding, we transform the MU-MIMO channel into parallel or approximately parallel SU-MIMO channels and the effective channel matrix for the $i$th user is
\begin{align}
\boldsymbol H_{{\rm eff}_i}=\boldsymbol H_i\boldsymbol P_i^a.
\end{align}
We perform the LR transformation on ${\boldsymbol H}_{{\rm
eff}_i}^T$ in the precoding scenario \cite{Windpassinger}, that is
\begin{align}
{\boldsymbol {\tilde H}}_{{\rm eff}_i}=\boldsymbol T_i\boldsymbol
H_{{\rm eff}_i}~{\rm and}~ \boldsymbol H_{{\rm eff}_i}=\boldsymbol T_i^{-1}{\boldsymbol {\tilde H}}_{{\rm eff}_i},
\end{align}
where $\boldsymbol T_i$ is a unimodular matrix with $\rm{det}|\boldsymbol
T_i|=1$ and all elements of $\boldsymbol T_i$ are complex
integers, i.e. $ t_{l,k} \in\mathbb{Z}+j\mathbb{Z}$. The physical meaning of the constraint $\rm{det}|\boldsymbol
T_i|=1$ is that the channel energy is unchanged after the LR transformation.

Following the LR transformation, we employ the linear precoding constraint to get the second precoding filter to parallelize each user's streams. The ZF precoding constraint is implemented for user $i$ as
\begin{align}
\boldsymbol {\tilde P}_{{\rm ZF}_i}^b={\boldsymbol {\tilde H}}_{{\rm
eff}_i}^H (\boldsymbol {\tilde H}_{{\rm eff}_i}{\boldsymbol {\tilde
H}}_{{\rm eff}_i}^H)^{-1}.
\end{align}
It is well-known that the performance of MMSE precoding is always better than that of ZF precoding.
We can get the second precoding filter by employing an MMSE precoding constraint.
The MMSE precoding is actually equivalent to the ZF precoding with respect to an extended
system model \cite{Wuebben, Habendorf}. The extended channel matrix
$\boldsymbol {\underline H}$ for the MMSE precoding scheme is defined as
\begin{align}
\boldsymbol {\underline H}=\begin {bmatrix}\boldsymbol H,\sqrt\alpha \boldsymbol I_{N_R}\end {bmatrix}.
\end{align}
By introducing the regularization factor $\alpha$, a trade-off between the level of MUI and the noise is introduced \cite{Michael}. Then, the MMSE precoding filter is obtained as
\begin{align}
\boldsymbol P_{\rm MMSE}=\boldsymbol A {\boldsymbol {\underline H}}^H(\boldsymbol
{\underline H}{\boldsymbol {\underline H}}^H)^{-1},
\end{align}
where $\boldsymbol A=\begin {bmatrix} \boldsymbol I_{N_T},\boldsymbol 0_{N_T \times N_R}\end {bmatrix}$, and the multiplication by $\boldsymbol A$ will not result in transmit power amplification since $\boldsymbol A\boldsymbol A^H=\boldsymbol I_{N_T}$.
From the mathematical expression in (33), the rows of $\boldsymbol {\underline H}$ determine the effective
transmit power amplification of the MMSE precoding. Correspondingly, the LR
transformation for the MMSE precoding should be applied to the transpose of the extended
channel matrix ${\boldsymbol {\underline H}}_{{\rm
eff}_i}^T={\begin {bmatrix}\boldsymbol H_{{\rm eff}_i},\sqrt\alpha\boldsymbol
I_{N_i}\end {bmatrix}}^T$, and the LR transformed
channel matrix ${\boldsymbol {\underline {\tilde H}}}_{{\rm eff}_i}$
is obtained as
\begin {align}
{\boldsymbol {\underline {\tilde H}}}_{{\rm eff}_i}=\boldsymbol {\underline T}_i\boldsymbol
{\underline H}_{{\rm eff}_i},
\end {align}
where $\boldsymbol {\underline T}_i$ is the unimodular matrix for $\boldsymbol {\underline H}_{{\rm eff}_i}$.
Then, the LR-aided MMSE precoding filter is given by
\begin{align}
 {\boldsymbol {\tilde P}}_{{\rm MMSE}_i}^b=\boldsymbol
 A_i {\boldsymbol {\tilde{\underline H}}}_{{\rm eff}_i}^H({\boldsymbol
{\tilde{\underline H}}}_{{\rm eff}_i}{\boldsymbol {\tilde{\underline
H}}}_{{\rm eff}_i}^H)^{-1},
\end{align}
where the matrix $\boldsymbol A_i=\begin {bmatrix} \boldsymbol I_{M_i},\boldsymbol 0_{M_i \times N_i}\end {bmatrix}$.
Finally, the combined second precoding matrix $\boldsymbol {\tilde P}^b$ for all users is
\begin{align}
\boldsymbol  {\tilde P}^b={\rm diag}\{\boldsymbol {\tilde P}_1^b,\boldsymbol  {\tilde P}_2^b,\ldots,\boldsymbol  {\tilde P}_K^b\}.
\end{align}
The overall precoding matrix is $\boldsymbol {\tilde P}=
\boldsymbol P^a \boldsymbol {\tilde P}^b$. Since the lattice reduced precoding matrix $\boldsymbol
{\tilde P}^b$ has near orthogonal columns, the required transmit power
will be reduced compared to the BD-type precoding algorithms. Thus, a better BER performance than that of the BD-type precoding algorithms can be achieved by the proposed LR-S-GMI-type precoding algorithms.

The received signal is finally obtained as
\begin{align}
\boldsymbol y=\boldsymbol H\boldsymbol {\tilde P}\boldsymbol d+\sqrt\gamma\boldsymbol n,
\end{align}
where $\gamma=\|\boldsymbol {\tilde P}\boldsymbol d\|^2$.
The main processing work left for the receiver is to quantize the received signal $\boldsymbol y$ to the nearest data vector and the decoding matrix $\boldsymbol G$ described in the BD-type \cite{Spencer02, Veljko}, QR/SVD-type \cite{Hua}, and GMI-type \cite{GMI} precoding algorithms is not needed anymore. The receiver structure is thus simplified, and a significant amount of transmit power can be saved which is very important considering the mobility of the distributed users.

The proposed precoding algorithms are called LR-S-GMI-ZF and LR-S-GMI-MMSE depending on the choice of the second precoding filter as given in (31) and (35), respectively. We will focus on the LR-S-GMI-MMSE precoding since a better performance is achieved. The implementing steps of the LR-S-GMI-MMSE precoding algorithm are summarized in Table II. By replacing the steps (8) and (9) in Table II with the formulation in (31), the LR-S-GMI-ZF precoding algorithm can be obtained. Similarly, the first precoding matrix can also be computed according to the GZI method in (28), and combined with (31) or (35) to get the second precoding matrix. Then, the LR-GZI-ZF or LR-GZI-MMSE precoding algorithms can be obtained, respectively.
\begin{table}[!t]
\caption{The LR-S-GMI-MMSE Precoding Algorithm} 
\centering 
\begin{tabular}{l l} 
\hline\hline 
Steps & Operations \\ [0.5ex] 
\hline 
& {\bf Applying the MMSE Channel Inversion}~~~~~~~~~~~~~~~~~~~~~\\
(1)& $\boldsymbol H^{\dag}_{{\rm mse}}  =(\boldsymbol H^H\boldsymbol
H+\alpha \boldsymbol I)^{-1}\boldsymbol H^H$\\
(2)& for i~=~1~:~$ K$\\
(3)&~~~~~$[\boldsymbol Q^\dag_{i,{\rm mse}}~ {\boldsymbol R^\dag_{i,{\rm mse}}}]={\rm QR}(\boldsymbol H^\dag_{i,{\rm mse}},~0)$\\
(4)&~~~~~$\boldsymbol P_i^a=\boldsymbol Q^\dag_{i,{\rm mse}}$\\
(5)&~~~~~$\boldsymbol H_{{\rm eff}_i}=\boldsymbol H_i\boldsymbol P_i^a$\\
(6)&~~~~~${\boldsymbol {\underline H}}_{{\rm
eff}_i}={\begin {bmatrix}\boldsymbol H_{{\rm eff}_i} ~ \sqrt\alpha\boldsymbol
I_{N_i}\end {bmatrix}}$\\
(7)&~~~~~$[\boldsymbol {\underline T}_i^T~
\boldsymbol{\underline H}_{{\rm eff}_i}^T]={\rm CLLL}({\boldsymbol {\underline {\tilde H}}}_{{\rm eff}_i}^T)$\\
(8)&~~~~~$\boldsymbol A_i=[\boldsymbol I_{M_i}~\boldsymbol 0_{M_i\times N_i}]$\\
(9)&~~~~~$ {\boldsymbol {\tilde P}}_{{\rm MMSE}_i}^b=\boldsymbol
 A_i {\boldsymbol {\tilde{\underline H}}}_{{\rm eff}_i}^H({\boldsymbol
{\tilde{\underline H}}}_{{\rm eff}_i}{\boldsymbol {\tilde{\underline
H}}}_{{\rm eff}_i}^H)^{-1}$\\
(10)&end\\
& {\bf Compute the overall precoding matrix}\\
(11)&$\boldsymbol P^a=[\boldsymbol P_1^a,~\boldsymbol P_2^a,~\ldots,~\boldsymbol P_K^a]$\\
(12)&$\boldsymbol  {\tilde P}^b={\rm diag}\{\boldsymbol {\tilde P}_1^b,\boldsymbol  {\tilde P}_2^b,\ldots,\boldsymbol  {\tilde P}_K^b\}$\\
(13)&$\boldsymbol  {\tilde P}=\boldsymbol P^a\boldsymbol  {\tilde P}^b$\\
& {\bf Calculate the scaling factor $\gamma$}\\
(14)&$\gamma=(\|\boldsymbol  {\tilde P}\boldsymbol d\|_F^2/E_s)$\\
& {\bf Get the received signal}\\
(15)& $\boldsymbol y=\boldsymbol H\boldsymbol {\tilde P}\boldsymbol d+\sqrt\gamma\boldsymbol n$\\
& {\bf Transform back from lattice space}\\
(16)&$\boldsymbol {\hat d}=\boldsymbol{\underline T}\lceil \boldsymbol y \rfloor$\\
[1ex] 
\hline 
\end{tabular}
\end{table}

\section {Performance Analysis}
In this section, we carry out an analysis of the performance of the proposed LR-S-GMI-type precoding algorithms.
We consider a performance analysis in terms of BER, sum-rate and computational complexity.
In the BER analysis part, we show that the residual interference matrix of the RBD precoding actually converges to an identity matrix, which is a new result in the literature so far. We also mathematically demonstrate that the residual interference of the proposed LR-S-GMI-type precoding algorithms converges to a zero matrix. Finally, we illustrate the quality of the effective channel matrices of the proposed and existing precoding algorithms by measuring their condition numbers. The maximum achievable sum-rate of the proposed LR-S-GMI-type precoding algorithms is given in the sum-rate analysis part. The computational complexity of the proposed and existing precoding algorithms is summarized in tables in the complexity analysis part. The trend of the computational complexity with the increase of the dimensions is also given and an analysis is developed.
\subsection{BER Performance Analysis}
For the BD precoding, the effective SU-MIMO channels are strictly parallel between each other after the first precoding filtering. For the RBD precoding, however, the residual interference $\boldsymbol{\overline H}_i{\boldsymbol P_i^a}^{\rm (RBD)}$ is not zero between the users. We use $\boldsymbol J_f$ to denote $\boldsymbol{\overline H}_i{\boldsymbol P_i^a}^{\rm (RBD)}$ for convenience. From (13), the following formula is obtained
\begin{align}
\boldsymbol J_f\boldsymbol J_f^H=\boldsymbol{\overline H}_i\boldsymbol{\overline V}_i(\boldsymbol{\overline \Sigma}_i^T\boldsymbol{\overline \Sigma}_i+\alpha\boldsymbol I_{N_T})^{-1}\boldsymbol{\overline V}_i^H{\boldsymbol{\overline H}_i}^H.
\end{align}
Mathematically, the quantity $\boldsymbol{\overline V}_i(\boldsymbol{\overline \Sigma}_i^T\boldsymbol{\overline \Sigma}_i+\alpha\boldsymbol I_{N_T})^{-1}\boldsymbol{\overline V}_i^H$ can be expressed as $(\boldsymbol{\overline H}_i^H\boldsymbol{\overline H}_i+\alpha\boldsymbol I_{N_T})^{-1}$. Substituting this into (38), the formula can be rewritten as
\begin{align}
\boldsymbol J_f\boldsymbol J_f^H=\boldsymbol{\overline H}_i(\boldsymbol{\overline H}_i^H\boldsymbol{\overline H}_i+\alpha\boldsymbol I_{N_T})^{-1}{\boldsymbol{\overline H}_i}^H.
\end{align}
With the increase of the SNR, $\alpha$ approaches zero and then we have
\begin{align}
\boldsymbol J_f\boldsymbol J_f^H\approx\boldsymbol{\overline H}_i(\boldsymbol{\overline H}_i^H\boldsymbol{\overline H}_i)^{-1}{\boldsymbol{\overline H}_i}^H.
\end{align}
By further manipulating the expression in (40), we obtain
\begin{align}
&\boldsymbol J_f\boldsymbol J_f^H\boldsymbol{\overline H}_i\approx\boldsymbol{\overline H}_i(\boldsymbol{\overline H}_i^H\boldsymbol{\overline H}_i)^{-1}{\boldsymbol{\overline H}_i}^H\boldsymbol{\overline H}_i=\boldsymbol{\overline H}_i\nonumber \\
&{\rm{Thus}}~~\boldsymbol J_f\boldsymbol J_f^H\approx\boldsymbol I_{N_T},
\end{align}
that is, the residual interference matrix $\boldsymbol J_f$ of the RBD precoding converges to an identity matrix at high SNRs. While, for the S-GMI precoding algorithm developed in Section IV with the SNR increase we have
\begin{align}
\boldsymbol{\overline H}_i{\boldsymbol P_i^a}=\boldsymbol{\overline H}_i\boldsymbol Q^\dag_{i,{\rm
mse}}\approx \boldsymbol 0.
\end{align}
By comparing (41) and (42), we can see that the impact of the residual interference of S-GMI precoding would be smaller than that of the conventional RBD precoding algorithm. Thus, we expect that a better BER performance is achieved by the S-GMI precoding algorithm over the conventional RBD precoding algorithm.

As pointed out in \cite{Peel}, the BER performance for a MIMO precoding system is actually determined by the energy of the transmitted signal $\gamma$. In order to reduce $\gamma$ and improve the BER performance further, we transform the effective channel $\boldsymbol H_{\rm eff}$ into the lattice space. By doing this, an improved basis ${\boldsymbol {\tilde H}}_{{\rm eff}}$ is computed. Actually, the LR transformed channel matrix ${\boldsymbol {\tilde H}}_{{\rm eff}}$ is quasi-orthogonal rather than strictly orthogonal. We can employ the condition number which is defined as \cite{MatrixTheo}
\begin{align}
{\rm cond}(\boldsymbol H)=\|\boldsymbol H\|_F\|\boldsymbol H^{-1}\|_F
\end{align}
to measure the orthogonality of the channel matrix.
From the above definition of the condition number in (43), we get that ${\rm cond}(\boldsymbol H)=1$ with equality for an orthogonal basis while matrices which are nearly singular have large condition numbers. In Fig. 2, the probability density functions (PDFs) of the condition numbers for the effective channel matrices are illustrated. For the effective channel matrix of the proposed LR-S-GMI-MMSE precoding algorithm, not only the spread in the condition numbers but also their average value is much smaller compared to the effective channel matrices achieved by the existing precoding algorithms. Therefore, a significant reduction in the required transmit power $\gamma$ is achieved and a better BER performance can be obtained by the proposed LR-S-GMI-MMSE precoding algorithm. Note for the special case of each user with a single receive antenna, the proposed LR-S-GMI-type precoding will not converge to GMI or S-GMI because the second precoding filter is designed in the lattice space.

\begin{figure}[htp]
\begin{center}
\def\epsfsize#1#2{0.95\columnwidth}
\epsfbox{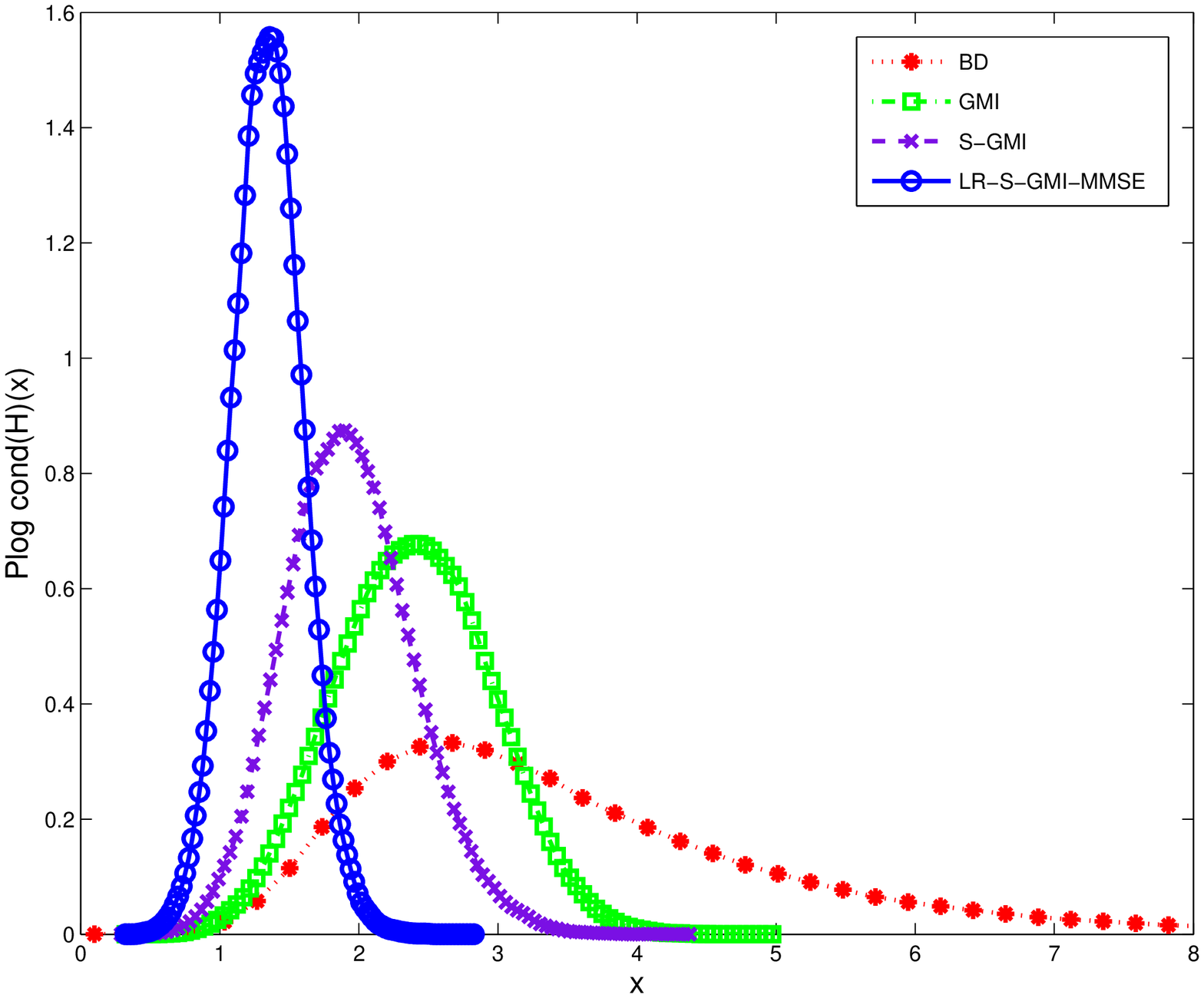} \vspace{-0.8em} \caption{PDFs of the natural
logarithm of ${\rm cond}(\boldsymbol H)$ for $6\times6$ matrices}
\end{center}
\end{figure}

\subsection{Achievable Sum-Rate Analysis}
Recall that at high SNRs, the MU-MIMO channel is approximately decoupled into equivalent SU-MIMO channels by applying the first precoding filtering in (23). Then, we can transform the MU-MIMO sum-rate analysis \cite{Vishwanath} to a set of SU-MIMO sum-rate analysis tasks. For the second precoding filter, the LR-aided MMSE precoding is actually equal to the LR-aided ZF precoding under the high SNR scenario. Therefore, the $i$th user's received signal is
\begin{align}
\boldsymbol y_i=\boldsymbol z_i+\sqrt\gamma_i\boldsymbol n_i,
\end{align}
where $\boldsymbol z_i=\boldsymbol T_i^{-1}\boldsymbol d_i$. By assuming that the average
transmit power is $\xi_i=1$, and because of the fact that $\boldsymbol H_{{\rm eff}_i}=\boldsymbol{U}_i\boldsymbol{\Sigma}_i{\boldsymbol{V}_i}^H$, we get the normalization factor $\gamma_i$ as
\begin{align}
\gamma_i &=\|\boldsymbol H_{{\rm eff}_i}^{-1}\boldsymbol z_i\|_{\rm F}^2=Tr(\boldsymbol{\Sigma}_i^{-2}\boldsymbol z_i\boldsymbol z_i^H) \nonumber \\
&={\sum _{l=1}^{L_{\rm eff}}{\xi_l^2\over\lambda_l^2}},
\end{align}
where the quantity $\lambda_l$ is the $l$th singular value of $\boldsymbol{\Sigma}_i$, and $\xi_l$ denotes the energy of the $l$th stream of $\boldsymbol z_i$.

From (45), the received SNR for the $l$th stream of user $i$ is obtained as
\begin{align}
{\rm SNR}_l={\xi_l^2\over {\sigma_n^2\sum _{m=1}^{L_{\rm eff}}{\xi_m^2\over\lambda_m^2}}}.
\end{align}
Then, the achievable sum-rate for user $i$ is given by
\begin {align}
C_i=\sum _{l=1}^{L_{\rm eff}}{\rm log}\biggl(1+ {\xi_l^2\over {\sigma_n^2\sum _{m=1}^{L_{\rm eff}}{\xi_m^2\over\lambda_m^2}}}\biggr)
=\sum _{l=1}^{L_{\rm eff}}{\rm log}\biggl(1+ {\xi_l^2\over {\sigma_n^2\gamma_i}}\biggr).
\end {align}
Note that the achievable sum-rate $C_i$ is degraded by the normalization factor $\gamma_i$. The value of $C_i$ approaches its maximum when ${\xi_1^2\over\lambda_1^2}={\xi_2^2\over\lambda_2^2}=\ldots={\xi_{L_{\rm eff}}^2\over\lambda_{L_{\rm eff}}^2}$, thus we have
\begin {align}
C_i\leq{\sum_{l=1}^{L_{\rm eff}}{\rm log}\biggl(1+ {\lambda_l^2\over {\sigma_n^2L_{\rm eff}}}}\biggr).
\end {align}
Finally, the maximum achievable sum-rate of the proposed LR-S-GMI-type precoding algorithms at high SNRs can be expressed as
\begin {align}
C=\sum_{i=1}^K{\sum_{l=1}^{L_{\rm eff}}{\rm log_2}\biggl(1+ {\lambda_l^2\over {\sigma_n^2L_{\rm eff}}}\biggr)}.
\end {align}
For the BD precoding, we multiply the decoding matrix $\boldsymbol G_i=\boldsymbol U_i^H$ at the $i$th user's receiver and the received signal is given by
\begin {align}
\boldsymbol y_i=\boldsymbol \Sigma_i\boldsymbol{\overline V}_i^{(0)}\boldsymbol{V}_i^{(1)}\boldsymbol d_i+ \boldsymbol U_i^H\boldsymbol n_i.
\end {align}
Due to the fact that the two precoding matrices $\boldsymbol{\overline V}_i^{(0)}$ and $\boldsymbol{V}_i^{(1)}$ are semi-unitary matrices, we get ${\boldsymbol{\overline V}_i^{(0)}}^H\boldsymbol{\overline V}_i^{(0)}=\boldsymbol I$ and ${\boldsymbol{V}_i^{(1)}}^H\boldsymbol{V}_i^{(1)}=\boldsymbol I$. Then, by applying the equivalence $Tr(\boldsymbol A\boldsymbol B\boldsymbol C)=Tr(\boldsymbol C\boldsymbol A\boldsymbol B)$, the normalization factor $\gamma_i^{(\rm BD)}$ for BD can be expressed as
\begin{align}
\gamma_i^{(\rm BD)} &=\|\boldsymbol{\overline V}_i^{(0)}\boldsymbol{V}_i^{(1)}\boldsymbol d_i\|^2=\|\boldsymbol d_i\|^2.
\end{align}
Since the statistical property of $\boldsymbol n_i$ is not changed by the multiplication with the unitary matrix $\boldsymbol U_i^H$, we get the $l$th received SNR as
\begin{align}
{\rm SNR}_l={\lambda_l^2\over {\sigma_n^2}}.
\end{align}
For simplicity, we do not consider the power loading between users and streams in the following derivation and
term this strategy as no power loading (NPL). Then, the achievable sum-rate for the BD precoding algorithm is given by
\begin {align}
C^{\rm (BD)}=\sum_{i=1}^K{\sum_{l=1}^{L_{\rm eff}}{\rm log_2}}\biggl(1+ {\lambda_l^2\over {\sigma_n^2}}\biggr).
\end {align}
By comparing the maximum achievable sum-rate of the proposed LR-S-GMI-type precoding algorithms in (49), we conclude that the sum-rate of the proposed LR-S-GMI-type precoding algorithms will be slightly inferior to that of the BD precoding algorithm at high SNRs. At low SNRs, however, we expect that the achieved sum-rate of the proposed LR-S-GMI-type precoding algorithms will be better than that of the BD precoding since a regularization factor is employed to mitigate the degradation by the noise term.

The sum-rate performance of the BD precoding is actually dependent on the power loading scheme being used. Hence, the BD precoding algorithm can achieve its maximum sum-rate performance by allocating the power between streams according to a WF power loading scheme. As pointed out in \cite{GMI}, we do not consider the power loading strategy for the RBD or the proposed LR-S-GMI-type precoding algorithms for two reasons. One is that it is not easy to identify the optimal power allocation coefficients because of the existence of residual interference. The second reason is that the MMSE condition (10) is already satisfied. Therefore, an allocation of different powers between streams is not needed.

\subsection{Computational Complexity Analysis}
In this section, we use the total number of floating point operations (FLOPs) to measure the
computational complexity of the precoding algorithms discussed above. It is worth noting that the lattice reduction algorithm has variable complexity, and the average complexity of the CLLL algorithm has been given in FLOPs by \cite{CLLL}.
A reduced and fixed complexity lattice reduction structure is proposed in \cite{FLLL}, however, we employ the conventional CLLL algorithm for the reason that the lattice reduction algorithm is not the main focus in this work.
The number of FLOPs for the complex QR decomposition and the real SVD operation are given in \cite{MatrixTheo}.
As shown in \cite{Zu03}, the number of FLOPs required by a $m\times n$ complex SVD operation is equivalent to its extended $2m\times 2n$ real matrix. The total number of FLOPs needed by the matrix operations is summarized below:
\begin {itemize}
\item {Multiplication of $m\times n$ and $n\times p$ complex matrices: $8mnp-2mp$;}

\item {QR decomposition of an $m\times n~(m\leq n)$ complex matrix: $16(n^2m-nm^2+{1\over 3}m^3)$;}

\item {SVD of an $m\times n ~ (m\leq n)$ complex matrix where only $\boldsymbol \Sigma$ and $\boldsymbol V$ are obtained: $32(nm^2+2m^3)$;}

\item {SVD of an $m\times n ~ (m\leq n)$ complex matrix where $\boldsymbol U$, $\boldsymbol \Sigma$ and $\boldsymbol V$ are obtained: $8(4n^2m+8nm^2+9m^3)$;}

\item {Inversion of an $m\times m$ real matrix using Gauss-Jordan elimination: $4m^3/3$.}
\end {itemize}

We illustrate the required FLOPs for the conventional RBD, S-GMI and LR-S-GMI-MMSE precoding algorithms in Table III, Table IV and Table V, respectively. The complexity of the QR/SVD RBD \cite{Hua} and LC-RBD-LR-MMSE precoding algorithms is already given in \cite{Zu03}. A system with $N_T=6$ transmit antennas and $K=3$ users each equipped with $N_i=2$ receive antennas is considered;
this scenario is denoted as the $(2,2,2)\times 6$ case.
For the $(2,2,2)\times 6$ case, the reduction in
the number of FLOPs obtained by the proposed LR-S-GMI-MMSE precoding is $73.6\%$, $69.5\%$, $59.1\%$ and $49.9\%$
as compared to the RBD, BD, QR/SVD RBD and LC-RBD-LR-MMSE precoding algorithms,
respectively. Clearly, the proposed LR-S-GMI-MMSE precoding algorithm requires the lowest complexity.

\begin{table}[htp]
\caption {Computational complexity of conventional RBD}
\centering 
\begin{tabular}{l l l l} 
\hline\hline 
Steps & Operations & Flops & Case\\
& & & $(2,2,2)\times 6$ \\ [0.5ex] 
\hline 
1 & $\boldsymbol{\overline U}_i\boldsymbol{\overline \Sigma}_i\boldsymbol{\overline V}_i^H$ & $32K(N_T\overline N_i^2+2\overline N_i^3 )$ & 21504\\ 
2 & $(\boldsymbol{\overline \Sigma}_i^T\boldsymbol{\overline \Sigma}_i+\alpha\boldsymbol I_{N_T})^{-{1\over2}}$ & $K(18N_T+\overline N_i)$ & 336\\ 
3 & $\boldsymbol{\overline V}_i\boldsymbol{\overline D}_i,(\boldsymbol{\overline D}_i \leftarrow 2)$ &$8KN_T^3$ & 5184\\ 
4 & $\boldsymbol H_i\boldsymbol P_i^a $ & $K(8N_i^2N_T-2N_i^2)$  & 552\\ 
5 & $\boldsymbol{U}_i\boldsymbol{\Sigma}_i{\boldsymbol{V}_i}^H$ &$64K({9\over 8}N_i^3+$ & 13248\\ 
& & $N_TN_i^2+{1\over 2}N_T^2N_i)$&Total 40824\\ [1ex] 
\hline 
\end{tabular}
\end{table}

\begin{table}[htp]
\caption {Computational complexity of S-GMI} 
\centering 
\begin{tabular}{l l l l} 
\hline\hline 
Steps & Operations & Flops & Case\\
& & & $(2,2,2)\times 6$ \\ [0.5ex] 
\hline 
1 & $\boldsymbol H^{\dag}_{{\rm mse}} $ & $({4\over 3}N_R^3+12N_R^2N_T$ &\\
& & $-2N_R^2-2N_RN_T)$&2736 \\
2 & $\boldsymbol Q_i^{\dag}\boldsymbol R_i^{\dag}$ & $16K(N_T^2N_i-N_T N_i^2+{1\over 3}N_i^3)$ &2432\\
3 & $\boldsymbol H_i\boldsymbol P_i^a $ & $K(8N_i^2N_T-2N_i^2)$ & 552 \\
4 & $\boldsymbol U_i\boldsymbol\Sigma_i\boldsymbol V_i^H$ &$64K({9\over 8}N_i^3+N_TN_i^2+$ & 13248\\
& & ${1\over 2}N_T^2N_i)$&Total 18968\\ [1ex]
\hline 
\end{tabular}
\end{table}

\begin{table}[htp]
\caption{Computational complexity of LR-S-GMI-MMSE} 
\centering 
\begin{tabular}{l l l l} 
\hline\hline 
Steps & Operations & Flops & Case\\
& & & $(2,2,2)\times 6$\\ [0.5ex] 
\hline 
1 & $\boldsymbol H^{\dag}_{{\rm mse}} $ & $({4\over 3}N_R^3+12N_R^2N_T$ &\\ 
& & $-2N_R^2-2N_RN_T)$&2736 \\
2 & $\boldsymbol Q_i^{\dag}\boldsymbol R_i^{\dag}$ & $16K(N_T^2N_i-N_T N_i^2+{1\over 3}N_i^3)$ &2432\\
3 & $\boldsymbol H_i\boldsymbol P_i^a $ & $K(8N_i^2N_T-2N_i^2)$ & 552 \\
4 & ${\boldsymbol {\underline {\tilde H}}}_{{\rm eff}_i}$ & CLLL &4787.58 \\ 
5 & ${\boldsymbol {\underline {\tilde H}}}_{{\rm eff}_i}^{\dag}$ & $K({4\over 3}N_i^3+12N_i^3-4N_i^2)$ &272\\
& & &Total 10780 \\
\hline 
\end{tabular}
\end{table}
In order to further reveal the relationship between the computational complexity and the system dimensions, we first fix the receive antenna configuration and assume that each user is equipped with $N_i=2$ antennas. Fig. 3 shows that with $N_i$ fixed, the computational complexity of the BD-type precoding algorithms grows relatively faster than the other precoding algorithms with the increase of $K$. The reason is that, the first SVD operation of the RBD precoding is implemented $K$ times on $\boldsymbol{\overline H}_i$ with dimension $\overline N_i\times N_T$. The QR decomposition of the LC-RBD-LR-type precoding also needs to be implemented $K$ times on $\boldsymbol{\overline H}_i$, but it requires less FLOPs because the QR decomposition is much simpler than the SVD operation in the case of same matrix dimensions \cite{MatrixTheo}. While, the S-GMI only needs one common channel inversion and the QR decomposition is computed on $\boldsymbol H_{i,{\rm mse}}$ with a lower dimension $N_i\times N_i$.
\begin{figure}[htp]
\begin{center}
\def\epsfsize#1#2{0.95\columnwidth}
\epsfbox{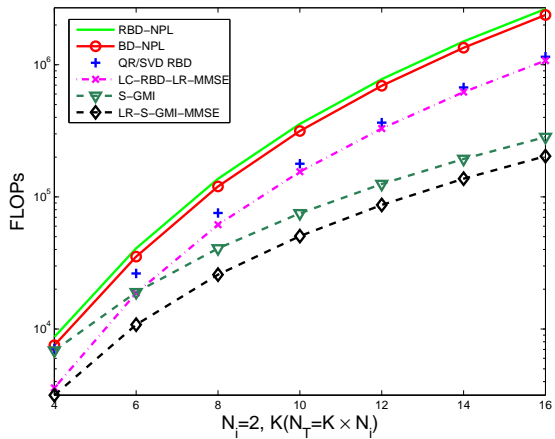} \vspace{-0.8em} \caption{Computational Complexity
- I Fixed $N_i$}
\end{center}
\end{figure}

Similarly, if we fix the number of users to $K=4$, the system dimensions will change with the variable $N_i$.
From Fig. 4, the S-GMI precoding algorithm can offer a much lower complexity than that of the BD, RBD and QR/SVD RBD precoding algorithms.
The reason is that, with $K$ fixed, the first $K$ SVD operations have a higher cost than the common channel inversion in (20) plus the QR decompositions in (24).
\begin{figure}[htp]
\begin{center}
\def\epsfsize#1#2{0.95\columnwidth}
\epsfbox{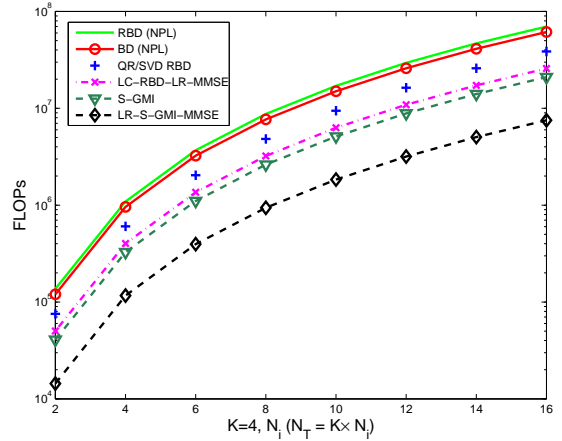} \vspace{-0.8em} \caption{Computational Complexity
- II Fixed $K$}
\end{center}
\end{figure}

The complexity of the proposed LR-S-GMI-MMSE precoding algorithm, however, shows the lowest computational complexity in Fig. 3 and Fig. 4. The reason is that we use a less complex LR transformation to replace the second SVD operation in S-GMI precoding algorithm. It is worth noting that with the increase of the system dimensions, the complexity reduction by the proposed LR-S-GMI-MMSE precoding algorithm becomes more considerable.

\section{Simulation Results}
A system with $N_T=8$ transmit antennas and $K=4$
users each equipped with $N_i=2$ receive antennas is considered;
this scenario is denoted as the $(2,2,2,2)\times 8$ case. The vector $\boldsymbol d_i$ of the $i$th user represents the data transmitted with QPSK modulation.
\subsection{Perfect Channel Scenario}
The channel matrix $\boldsymbol H_i$ of the $i$th user is modeled as a complex Gaussian channel matrix with zero mean and unit variance. We assume an uncorrelated block fading channel, that is, the channel is static during each transmit
packet and there is no correlation between the antennas. We also
assume that the channel estimation is perfect at the receive side
and the feedback channel is error free. The number of simulation
trials is $10^6$ and the packet length is $10^2$ symbols.
The $E_b/N_0$ is defined as $E_b/N_0={N_R\xi\over N_TM\sigma_n^2}$ with $M$ being the number of transmitted information bits per channel symbol.

Fig. 5 shows the BER performance of the proposed and existing
precoding algorithms. The GMI, QR/SVD RBD and RBD precoding algorithms share the same BER performance.
A better BER performance is achieved by the proposed S-GMI precoding scheme compared to the BD, GMI, QR/SVD RBD, and RBD precoding algorithms.
The reason is that the residual interference between the users can be suppressed further by the S-GMI precoding scheme as we analyzed in Section V.A.
The proposed LR-S-GMI-MMSE precoding algorithm shows the best BER performance.
At the BER of $10^{-2}$, the LR-S-GMI-MMSE precoding has
a gain of more than 5.5 dB compared to the RBD precoding. It is worth noting that the BER performance of the RBD precoding is
outperformed by the proposed LR-S-GMI-MMSE precoding in the whole $E_b/N_0$ range and gains become more significant with the increase of $E_b/N_0$.
From the analysis developed in Section V.A, a better channel quality as measured by the condition number of the effective channel is achieved by the proposed LR-S-GMI-MMSE precoding. Therefore, the required transmit power $\gamma$ is reduced and a better BER performance is obtained.
\begin{figure}[htp]
\begin{center}
\def\epsfsize#1#2{0.95\columnwidth}
\epsfbox{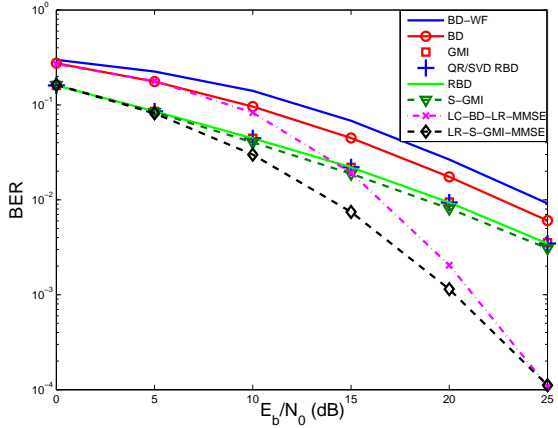} \vspace{-0.8em} \caption{BER performance,
$(2,2,2,2)\times 8$ MU-MIMO}
\end{center}
\end{figure}

Fig. 6 illustrates the sum-rate of the proposed and existing precoding
algorithms. The sum-rate is calculated using \cite{Vishwanath}:
\begin{align}
 C={\rm log}({\rm det}(\boldsymbol I+\sigma_n^{-2}\boldsymbol H\boldsymbol P \boldsymbol P^H \boldsymbol H^H)) ~~{\rm (bits/Hz)}.
\end{align}
The BD precoding with WF power loading shows a better sum-rate performance than the BD precoding without power loading as we mentioned in Section V.B. However, as shown in Fig. 5, the BER performance is degraded by applying this WF scheme. Similar to the BER figure, the GMI, QR/SVD RBD, and RBD precoding algorithms show a comparable sum-rate performance. The S-GMI precoding also achieves the sum-rate performance of the RBD precoding.
The proposed LR-S-GMI-MMSE precoding algorithm illustrates almost the same sum-rate performance as the RBD precoding at low $E_b/N_0$s. At high $E_b/N_0$s, however, the sum-rate of LR-S-GMI-MMSE precoding is slightly inferior to that of the RBD precoding and approaches the performance of the BD precoding as we analyzed in Section~V.B.

\begin{figure}[htp]
\begin{center}
\def\epsfsize#1#2{0.95\columnwidth}
\epsfbox{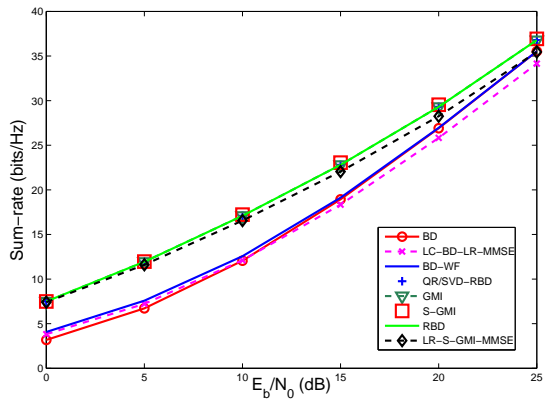} \vspace{-0.8em} \caption{Sum-rate performance,
$(2,2,2,2)\times 8$ MU-MIMO}
\end{center}
\end{figure}

\subsection{The impact of imperfect channels}
The use of perfect CSI is impractical in wireless systems due to the often inaccurate channel estimation and the CSI feedback errors. From \cite{Zu02, Windpassinger02}, the estimation errors or feedback errors can be modeled as a complex random Gaussian noise $\boldsymbol E$ with i.i.d. entries of zero mean and variance $\sigma_e^2$. Another factor that usually needs to be considered in the MU-MIMO systems is the antenna correlation at the transmit side \cite{Michel}. In this work, we simulate the correlated channel based on the exponential correlation model in \cite{Sergey}.
The imperfect channel matrix $\boldsymbol H_e$ is defined as
\begin{align}
\boldsymbol H_e=\boldsymbol H\boldsymbol R_T^{1\over 2}+\boldsymbol E,
\end{align}
where the quantity $\boldsymbol R_T$ is a transmit covariance matrix with the elements defined below
\begin{align}
R_{ij}=\left\{\begin{array}{ll} r^{j-i},& i\leq j\\ r_{ji}^*,& i>j \end{array} \right.,|r|\leq 1
\end{align}
where $r$ is the correlation coefficient between any two neighboring antennas.
The precoding matrix $\boldsymbol P$ has to be designed based on the imperfect channel $\boldsymbol H_e$ while the physical channel is $\boldsymbol H$ during each transmission.

Fig. 7 gives the BER performance of the above precoding algorithms under $\boldsymbol H_e$ with $|r|=0.2$ at $E_b/N_0=15~\rm dB$. All the above precoding algorithms are affected by the imperfect channel $\boldsymbol H_e$.
The proposed LR-S-GMI-MMSE precoding algorithm outperforms the RBD precoding algorithm when $\sigma_e^2$ is below $0.12$, however, the performance of the RBD precoding algorithm decays more slowly.
\begin{figure}[htp]
\begin{center}
\def\epsfsize#1#2{0.95\columnwidth}
\epsfbox{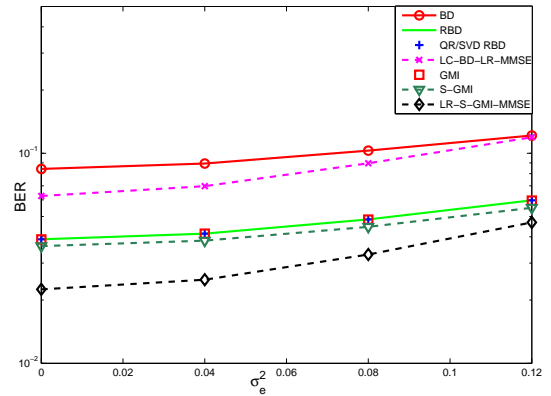} \vspace{-0.8em} \caption{BER as a function of
$\sigma_e^2$ for a fixed $E_b/N_0$=15 dB}
\end{center}
\end{figure}

\section{conclusion}
Based on a channel inversion technique, low-complexity high-performance LR-S-GMI-type precoding algorithms have been proposed for MU-MIMO systems in this paper. Compared to the BD-type precoding algorithms, the complexity of the precoding process
is reduced and a considerable BER gain is achieved by the
proposed algorithms at a cost of a slight sum-rate loss at high SNRs. The BER performance, the achievable sum-rate and the computational complexity of the LR-S-GMI-type precoding algorithms have been analyzed and compared to existing precoding algorithms. Since the LR-S-GMI-type precoding algorithms are only implemented at the transmit side, the decoding matrix is not needed any more at the receive side compared to the BD-type precoding algorithms. Then, the structure of the receiver can be simplified, which is an additional benefit from the proposed LR-S-GMI-type precoding algorithms. The proposed algorithms also show a robust performance in the presence of imperfect CSI and spatial correlation, which emphasizes their value for practical applications.


\begin{thebibliography}{30}
{\footnotesize
\bibitem{Zu01}
K. Zu, R. C. de Lamare and M. Haardt, "Low-complexity lattice reduction-aided channel inversion methods for large multi-user MIMO systems", in \textit{Proc. IEEE Asilomar Conference on Signals, Systems and Computers}, Pacific Grove, USA, Oct. 2012.

\bibitem{Paulraj01}
A. Paulraj, R. Nabar and D. Gore, \textit{Introduction to Space-Time Wireless Communications}. Cambridge University Press, 2003.

\bibitem{Tse}
D. Tse and P. Viswanath, \textit{Fundamentals of Wireless Communications}. Cambridge University Press, 2005.

\bibitem{Lte}
\textit{Requirements for Further Advancements for E-UTRA (LTE-Advanced)}, 3GPP TR 36.913 Standard, 2011.

\bibitem{80211ac}
\textit{Wireless LAN Medium Access Control (MAC) and Physical Layer (PHY) Specifications: Enhancements for Very High Throughput for Operation in
Bands Below 6GHz}, IEEE P802.11ac/D1.0 Stdandard., Jan. 2011.

\bibitem{chiu}
C. Chiu, J. Yan and R. Murch, "24-Port and 36-Port antenna
cubes suitable for MIMO wireless communications," \textit{IEEE
Trans. Antennas Propag.}, vol.56, no.4,
pp. 1170-1176, Apr. 2008.

\bibitem{Fredrik}
F. Rusek, D. Persson, B. Lau, E. Larsson, T. Marzetta, O. Edfors and F. Tufvesson, "Scaling up MIMO: opportunities and challenges with very large arrays,"
\textit{IEEE Signal Processing Mag. (1991-present)}, to be published.

\bibitem{Michael}
M. Joham, W. Utschick and J. A. Nossek, "Linear transmit processing
in MIMO communications systems," \textit{IEEE Trans. Signal Process.}, vol. 53  no. 8, pp. 2700–2712, Aug. 2005.

\bibitem{Cai1}
Y. Cai, R. C. de Lamare and D. L. Ruyet, "Transmit Processing
Techniques Based on Switched Interleaving and Limited Feedback for
Interference Mitigation in Multiantenna MC-CDMA Systems,"
\textit{IEEE Transactions on Vechicular Technology}, vol. 60, no. 4,
May 2011.

\bibitem{Cai2}
Y. Cai, R. C. de Lamare, R. Fa, "Switched interleaving techniques
with limited feedback for interference mitigation in DS-CDMA
systems," \textit{IEEE Transactions on Communications}, vol. 59, no.
7, Jul. 2011.


\bibitem{Spencer02}
Q. H. Spencer, A. L. Swindlehurst and M. Haardt, "Zero-forcing
methods for downlink spatial multiplexing in multiuser MIMO
channels," \textit{IEEE Trans. Signal Process.}, vol. 52, no. 2,
pp. 461-471, Feb. 2004.

\bibitem{Choi}
L. U. Choi and R. D. Murch, "A transmit preprocessing technique for
multiuser MIMO systems using a decomposition approach," \textit{IEEE
Trans. Wireless Commun.}, vol. 3, no. 1, pp. 20-24, Jan. 2004.

\bibitem{Veljko}
V. Stankovic and M. Haardt, "Generalized design of multi-user MIMO precoding matrices,"
\textit{IEEE Trans. Wireless Commun.}, vol. 7, no. 3, pp. 953-961, Mar. 2008.

\bibitem{Chae}
C. B. Chae, S. Shim and R. W. Heath, "Block diagonalized vector
perturbation for multiuser MIMO systems," \textit{IEEE Trans.
Wireless Commun.}, vol. 7, no. 11, pp. 4051-4057, Nov. 2008.

\bibitem{GMI}
H. Sung, S. Lee and I. Lee, "Generalized channel inversion methods for multiuser MIMO systems,"
\textit{IEEE Trans. Commun.}, vol. 57, no. 11, pp. 3489-3409, Nov. 2009.

\bibitem{Hua}
H. Wang, L. Li, L. Song and X. Gao, "A linear precoding scheme for
downlink multiuser MIMO precoding systems," \textit{IEEE Commun. Lett.}, vol. 15, no. 6, pp. 653–655, Jun. 2011.

\bibitem{MatrixTheo}
G. Golub and C. V. Loan, \textit{Matrix Computaitons}. The Johns Hopkins
University Press, 1996.

\bibitem{LR}
D. Wübben, D. Seethaler, J. Jaldén and G. Matz, "Lattice reduction: a survey with applications in wireless communications," \textit{IEEE Signal Processing Mag. (1991-present)}, vol. 28, no. 3, pp. 70-91,  May 2011.

\bibitem{Zu03}
K. Zu and R. C. de Lamare, "Low-complexity lattice reduction-aided regularized block diagonalization for MU-MIMO systems," \textit{IEEE Commun. Lett.}, vol. 16, no. 6, pp. 925-928, Jun. 2012.

\bibitem{Zu02}
K. Zu, R. C. de Lamare and M. Haardt, "Lattice reduction-aided regularized block diagonalization for multiuser MIMO systems", \textit{Proc. 2012 IEEE Wireless Communications and Networking Conf. (WCNC)}, Paris, France, Apr. 2012, pp. 131-135.

\bibitem{Peel}
C. B. Peel, B. M. Hochwald and A. L. Swindlehurst, "A vector-perturbation technique for near capacity multiantenna multiuser communication - Part I: Channel inversion and regularization," \textit{IEEE Trans. Commun.}, vol. 52, no. 1, pp. 195–202, Jan. 2005.

\bibitem{Hochwald}
B. M. Hochwald, C. B. Peel and A. L. Swindlehurst, "A vector-perturbation technique for near capacity multiantenna multiuser communication - Part II: Perturbation," \textit{IEEE Trans. Commun.}, vol. 53, no. 3, pp. 537–544, Mar. 2005.

\bibitem{LLL}
A. K. Lenstra, H. W. Lenstra and L. Lov\'asz, "Factoring polynomials with rational coefficients," \textit{Math. Ann}, vol. 261, pp. 515-534, 1982.

\bibitem{delamarespl07} R. C. de Lamare and R.
Sampaio-Neto, ``Reduced-Rank Adaptive Filtering Based on Joint
Iterative Optimization of Adaptive Filters", \textit{IEEE Signal
Processing Letters}, Vol. 14, no. 12, December 2007.

\bibitem{delamaretvt10}
R. C. de Lamare and R. Sampaio-Neto, ``Reduced-Rank Space-Time
Adaptive Interference Suppression With Joint Iterative Least Squares
Algorithms for Spread-Spectrum Systems," \textit{IEEE Transactions
on Vehicular Technology}, vol.59, no.3, March 2010, pp.1217-1228.


\bibitem{jidf}
R. C. de Lamare and R. Sampaio-Neto, ``Adaptive Reduced-Rank
Processing Based on Joint and Iterative Interpolation, Decimation,
and Filtering," \textit{IEEE Trans. Sig. Proc.}, vol. 57, no. 7,
July 2009, pp. 2503 - 2514.

\bibitem{CLLL}
Y. H. Gan, C. Ling and W. H. Mow, "Complex lattice reduction
algorithm for low-complexity full-diversity MIMO detection," \textit{IEEE
Trans. Signal Process.}, vol. 57, no. 7, pp. 2701-2710, Jul. 2009.

\bibitem{Windpassinger}
C. Windpassinger and R. Fischer, "Low-complexity near-maximum
likelihood detection and precoding for MIMO systems using lattice
reduction," in \textit{Proc. IEEE Information Theory Workshop}, Paris, France, Mar. 2003, pp. 345-348.

\bibitem{Wuebben}
D. Wübben, R. Böhnke, V. Kühn and K. Kammeyer,
"Near-maximum-likelihood detection of MIMO systems using MMSE-based
lattice-reduction," in \textit{Proc. IEEE International Conference on
Communications (ICC)}, Paris, France, Jun. 2004, pp. 798-802.

\bibitem{spadf} R. C. de Lamare
and R. Sampaio-Neto, "Minimum Mean Squared Error Iterative
Successive Parallel Arbitrated Decision Feedback Detectors for
DS-CDMA Systems," \emph{IEEE Transactions on Communications.} vol.
56, no. 5, May, 2008.

\bibitem{delamaretvt}
R.C. de Lamare and R. Sampaio-Neto, ``Adaptive Reduced-Rank
Equalization Algorithms Based on Alternating Optimization Design
Techniques for MIMO Systems," \textit{IEEE Trans. Vehicular
Technology}, vol. 60, no. 6, pp.2482-2494, July 2011.

\bibitem{Habendorf}
R. Habendorf and G. Fettweis, "On ordering optimization for MIMO systems with decentralized receivers," in \textit{Proc. 63rd IEEE Vehicular Technology Conference(VTC)}, Melbourne, Australia, May 2006, pp. 1844-1848.

\bibitem{LrDiversity}
M. Taherzadeh, A. Mobasher and A. K. Khandani, "LLL reduction achieves the receive diversity in MIMO decoding," \textit{IEEE Trans. Inf. Theory}, vol. 53, no. 12, pp. 4801–4805, Dec. 2007.

\bibitem{Vishwanath}
S. Vishwanath, N. Jindal and A. J. Goldsmith, "On the capacity of
multiple input multiple output broadcast channels," in \textit{Proc. IEEE
International Conference on Communications (ICC)},
New York, USA, Apr. 2002, pp. 1444-1450.

\bibitem{FLLL}
H. Vetter, V. Ponnampalam, M. Sandell and P. A. Hoeher, "Fixed complexity LLL algorithm," \textit{IEEE Trans. Signal Process.}, vol. 57, no. 4, pp. 1634–1637, Apr. 2009.

\bibitem{Windpassinger02}
C. Windpassinger, "Detection and precoding for multiple input multiple output channels," Ph.D. dissertation, Univ. Erlangen-Nurnberg, Erlangen, Germany, 2004.

\bibitem{Michel}
M. T. Ivrlac, W. Utschick and J. A. Nossek, "Fading correlations in wireless MIMO communicaitons systems," \textit{IEEE J. Sel. Areas Commun.}, vol. 21, no. 5, pp. 819–828, Jun. 2003.

\bibitem{Sergey}
S. L. Loyka, "Channel capacity of MIMO architecture using the exponential correlation matrix," \textit{IEEE Commun. Lett.}, vol. 5, no. 9, pp. 369–371, Sep. 2001.

}
\end{thebibliography}
\end{document}